\documentclass{aa}

\usepackage{graphics}
\usepackage{epsfig}

\def\bib{\bibitem{}}

\newcommand{\xia}{\overline{\xi}}

\newcommand{\rhob}{\overline{\rho}}
\newcommand{\gam}{\gamma}

\newcommand{\beq}{\begin{equation}}
\newcommand{\eeq}{\end{equation}}
\newcommand{\lag}{\langle}
\newcommand{\rag}{\rangle}
\newcommand{\Om}{\Omega_{\rm m}}
\newcommand{\Ol}{\Omega_{\Lambda}}
\begin{document}
%
%
%
%
\renewcommand{\textfraction}{.01}
\renewcommand{\topfraction}{0.99}
\renewcommand{\bottomfraction}{0.99}
\setlength{\textfloatsep}{2.5ex}
\thesaurus{Sect.02 (12.12.1; 11.05.2; 11.17.3; 11.09.3; 11.17.1; 11.03.1)} 
\title{The redshift evolution of bias and baryonic matter distribution.}
\author{Patrick Valageas\inst{1}, Joseph Silk\inst{2,}\inst{3} 
\and Richard Schaeffer\inst{1}}
\institute{Service de Physique Th\'eorique, CEA Saclay, 91191 
Gif-sur-Yvette, France
\and
Astrophysics, Department of Physics, Keble Road, Oxford OX1 3RH, U.K.
\and
Institut d'Astrophysique de Paris, CNRS, 98bis Boulevard Arago,
F-75014 Paris, France}
\date{Received ;  }
\date{Received / Accepted }
\maketitle 
\markboth{P. Valageas, J. Silk and R. Schaeffer: The redshift evolution 
of bias and baryonic matter distribution.}  {P. Valageas, J. Silk and
R. Schaeffer: The redshift evolution of bias and baryonic matter
distribution.}

\begin{abstract}

We study the distribution of baryonic and luminous matter within the
framework of a hierarchical scenario. Using an analytical model for
structure formation which has already been checked against
observations for galaxies, Lyman-$\alpha$ clouds, clusters and reionization
processes, we present its predictions for the bias of these
objects. We describe its dependence on the luminosity (for galaxies or
quasars) or the column density (for Lyman-$\alpha$ absorbers) of the
considered objects. We also study its redshift evolution, which can
exhibit an intricate behaviour. These astrophysical objects
do not trace the dark matter
density field, the Lyman-$\alpha$
 forest clouds being undercorrelated and the bright galaxies overcorrelated, 
while the intermediate class of
 Lyman-limit systems is seen to sample the matter field quite well. 
 We also present the distribution of baryonic matter
over these various objects. We show that light does not trace baryonic
mass, since bright galaxies which contain most of the stars only form
a small fraction of the mass associated with virialized and cooled
halos. We consider two cosmologies: a critical density universe and an open
universe. In both cases, our results agree with observations and show
that hierarchical scenarios provide a good model for structure
formation and can describe a wide range of objects which spans at least the 
seven orders of magnitude in mass for which data exist. More detailed 
observations, in particular of the clustering
evolution
of galaxies, will constrain the astrophysical models involved.

\end{abstract}


\section{Introduction}

In the standard cosmological scenario large-scale structures observed
in the present universe were formed by the amplification through
gravitational instability of small primordial density fluctuations
(usually assumed to be gaussian). Within hierarchical models (as in
the CDM case: Peebles 1982) small scales collapse first to form bound
objects which later merge to build more massive halos as larger scales
become non-linear. This describes the evolution under the action of
gravity of the dark matter component. However, observed objects
(e.g., quasars, galaxies, Lyman-$\alpha$ clouds...) are formed by
baryonic matter (gas, stars) which does not necessarily follow the
behaviour of the dark matter component since it undergoes additional
processes (e.g., radiative cooling). Moreover, by looking at specific
astrophysical objects (e.g. bright galaxies) one selects peculiar dark
matter environments where the former are most likely to arise. Thus,
to relate observations to the underlying dark matter component (which
is usually gravitationally dominant) and to test scenarios for
structure formation one needs to study the link between the dark and
baryonic components. This can also constrain the astrophysical models
used to build galaxies and other objects from the density field.

One tool to describe the distribution of objects in the universe, in
addition to their abundance, is their two-point correlation function
which measures their clustering. Then one often defines the bias $b^2$
of these halos as the ratio of this quantity to the dark matter
correlation function, which shows in a clear quantitative way
whether these objects are more or less clustered than the dark matter
component. In this article, we present the bias (and the correlation
length) associated with various objects (galaxies, quasars,
Lyman-$\alpha$ clouds, clusters). The formulation of this bias is obtained 
(Schaeffer 1985,1987; Bernardeau \& Schaeffer 1992,1999)
within the framework of a general description
of structure formation in the universe provided
by a {\it  scaling model}  (Schaeffer 1984; Balian \& Schaeffer 1989)
which describes the density field in the non-linear regime. These predictions specifically take into account the deep non-linearity of the density field and are conceptually  different from those obtained using linear theory (e.g., Mo \& White 1996). Indeed, the latter authors evaluate the bias 
of primordial overdensities that will collapse later, using
linear theory, with the implicit assumption that this bias is conserved during the course of evolution. They find the bias depends separately on the mass, radius and overdensity.
 Here, we use the bias calculated directly using non-linear theory. This bias is seen to depend on very specific internal properties of the considered non-linear objects, namely on a unique parameter $x$, quite close to the internal velocity dispersion within the objects, thus exhibiting a scaling law. Such a specific dependence has indeed been found to hold in numerical simulations (Munshi et al. 1999b), and we consider this approach to provide a more trustworthy description of the present clustering properties as well as their evolution at higher redshifts.

In the present paper, we specifically study the dependence of the
bias on the observable properties of the considered objects (luminosity, column
density) and its evolution with redshift. In contrast with most
previous works, our analytical model relating dark to observable mass is not an 
ad-hoc
parameterization. Indeed, it is simply a consequence of a general
model (Valageas \& Schaeffer 1997) which has already been applied to galaxies 
(Valageas \&
Schaeffer 1999), Lyman-$\alpha$ clouds (Valageas et al. 1999), clusters 
(Valageas \& Schaeffer 2000) and
reionization studies (Valageas \& Silk 1999a,b). Thus, its predictions
have already been compared with many observations (e.g., galaxy
luminosity function, column density distribution of Lyman-$\alpha$
clouds, amplitude of the UV background radiation field, cluster X-ray luminosity function,...). Hence there is no free parameter chosen in this article. Next, in addition to the bias we also present the distribution of baryonic matter over
the various objects we describe.

This article is organized as follows. In Sect.\ref{Multiplicity
functions} we describe our prescription for mass functions, while in
Sect.\ref{Bias} we recall how the bias is obtained in this model and we compare our formalism with other prescriptions. In
Sect.\ref{Critical universe} we present our results for the case of a
critical density universe. We first describe the clustering of galaxies,
quasars, Lyman-$\alpha$ clouds and clusters and then we study the repartition of
baryonic matter within various classes of objects. Finally, in
Sect.\ref{Open universe} we present the case of an open universe.

\section{Multiplicity functions}
\label{Multiplicity functions}

Since we wish to describe the properties of various classes of objects
like Lyman-$\alpha$ clouds and galaxies which are defined by specific
constraints we need a formalism which can handle more general
multiplicity functions than the usual mass function of
``just-virialized halos''. To this order, we shall assume that the
non-linear density field obeys the scaling model detailed in Balian \&
Schaeffer (1989). Thus, we define each class
of objects (Lyman-$\alpha$ clouds or galaxies) by a constraint
$\Delta(M,z)$ on the density contrast of the underlying dark-matter
halo of mass $M$ at redshift $z$. The ``just-virialized halos''
correspond to the special case $\Delta(M,z) = \Delta_c(z)$ with
$\Delta_c \sim 177$. For our purposes we consider the cases of a
constant density contrast, arising from a virialization constraint,
and of a constant radius (i.e. $(1+\Delta) \propto M$), arising from a
fixed cooling radius or Jeans length. In any case, we write the
multiplicity function of these objects as (see Valageas \& Schaeffer
1997 for details):
\beq
\eta(M,z) \frac{dM}{M}  = \frac{\rhob}{M} \; x^2 H(x) \; \frac{dx}{x}  \; .  
\label{etah}
\eeq
Here, the parameter $x$ associated with a halo of mass $M$ at redshift
$z$ is defined by:
\beq
x(M,z) =  \frac{1+\Delta(M,z)}{\xia[R(M,z),z]} \; ,
\label{xnl}
\eeq
where 
\[
\xia(R) =   \int_V \frac{d^3r_1 \; d^3r_2}{V^2} \; \xi_2 ({\bf r}_1,{\bf r}_2) 
 \;\;\;\;\; \mbox{with} \;\;\;\;\; V= \frac{4}{3} \pi R^3
\]
is the average of the two-body correlation function $\xi_2 ({\bf
r}_1,{\bf r}_2)$ over a spherical cell of radius $R$ and provides the
measure of the density fluctuations in such a cell (thus typical objects have $x 
\sim 1$). The scaling
function $H(x)$ depends on the initial power-spectrum of the density
fluctuations and must be taken from numerical simulations. However,
from theoretical considerations (Balian \& Schaeffer 1989) it is
expected to satisfy the asymptotic behaviour:
\[
x \ll 1 \; : \; H(x) \propto x^{\omega-2} \hspace{0.3cm} ,
\hspace{0.3cm} x \gg 1 \; : \; H(x) \propto x^{\omega_s-1} \;
e^{-x/x_*}
\]
with $\omega \simeq 0.5$, $\omega_s \sim -3/2$, $x_* \sim 10$ to 20,
and by definition it must satisfy
\beq
\int_0^{\infty}  x \; H(x) \; dx   =  1 \; .
\eeq
These properties have been checked for various $P(k)$ by Bouchet et al (1991),
Colombi et al. (1992,1994,1995,1997), Munshi et al. (1999a),
 while the mass functions obtained from
(\ref{etah}) for various constraints $\Delta(M)$ (for a constant $\Delta$
which was taken from $-0.5$ to $5000$ or for $(1+\Delta) \propto M$) have
been shown to provide reasonable approximations to numerical results
in Valageas et al. (2000).

\section{Bias}
\label{Bias}

\subsection{Our formulation}
\label{Our formulation}

Within the framework of the scale-invariance of the many-body matter
correlation functions $\xi_p({\bf r}_1, ..., {\bf r}_p)$ which led to
the mass function (\ref{etah}), one can show that in the highly
non-linear regime the bias characteristic of two objects factorizes
and is a function of the sole parameter $x$ introduced in
Sect.\ref{Multiplicity functions}, see Bernardeau \& Schaeffer
(1992,1999). Thus we write the correlation function of objects of type (1)
with objects of type (2) as: 
\beq 
\xi_{1,2}(r) = b(x_1) \; b(x_2) \; \xi(r)
\label{bx}
\eeq
with
\beq
x \ll 1 : \; b(x) \propto x^{(1-\omega)/2} \hspace{0.5cm} \mbox{and} 
\hspace{0.5cm} x \gg 1 : \; b(x) \propto x \; .
\label{bx1}
\eeq
This behaviour has been sucessfully checked in  numerical simulations by Munshi et al. (1999c). This allows us to obtain the redshift evolution of the 
bias associated with galaxies, Lyman-$\alpha$ clouds or quasars, once we define the constraints $\Delta(M,z)$ which characterize these various astrophysical objects (that in turn define the proper value of $x$ to be used).

\subsection{Comparison with some other models}
\label{Comparison with some other models}

We can note that alternative models have been proposed in the litterature to describe the clustering of galaxies. Here we briefly compare our prescription to such models, in order to clarify their main differences. Thus, semi-analytic models of galaxy and dark matter clustering based on Press-Schechter-like ideas have recently been proposed (Seljak 2000; Peacock \& Smith 2000). They describe the dark matter density field as a collection of smooth halos with a mean density profile. The mass function of these ``just-virialized'' halos (defined by an overall density contrast $\Delta=\Delta_c \sim 177$) is obtained from a slightly modified Press-Schechter mass function (Press \& Schechter 1974). By construction, such a model neglects the substructures of virialized halos since they are described by an average density profile. This is not a problem at large scales where the correlations among galaxies are governed by the distribution of their host virialized halos (note that since the Press-Schechter prescription is based on the linear density field these correlations should actually apply to very large scales which are still in the linear regime). However, at scales of the order of the size of large ``just-virialized'' halos the clustering of galaxies is set by the distribution of these galaxies within those larger objects (which actually correspond to clusters of galaxies at large masses).

Hence one needs to add a specific prescription to cope with this regime. Thus, one introduces the mean number $\lag N \rag(M)$ of galaxies contained in halos of mass $M$. This function is derived from N-body simulations or observations. Then, one puts a galaxy at the center of the halo and the $(N-1)$ remaining galaxies at random within the halo, following the mean dark matter density profile. In order to include the fluctuations of $N(M)$ one can even reproduce the second moment $\lag N (N-1)\rag(M)$ obtained from simulations (Seljak 2000). However, it is clear that the predictive power of such a model is quite limited. Indeed, the output of the model (i.e. the two-point correlation function of galaxies) is almost directly related to the input (the first two moments of the galaxy distribution within virialized halos). Thus, this can at best serve as a test of N-body simulations if this data is taken from simulations (while it would not make much sense to obtain this data from the observations one would later on try to recover). In particular, in order to derive the $p-$point correlation function of galaxies in this line one would need the moment $\lag N^p \rag$ of order $p$... On the other hand, if one only includes the mean $\lag N \rag(M)$ (taken from observations for instance) as input, one can argue that the model tests whether the spatial galaxy distribution is well described by the mean dark matter density profile. However, as shown in Valageas (1999) such a model contradicts the results of numerical simulations for the dark matter distribution itself when one takes into account the constraints provided by the first five order moments of counts-in-cells statistics. This implies that one could not get higher-order moments (e.g., the three-point galaxy correlation function) from this model.

Moreover, it is not clear how much information one can draw from the mere comparison of the galaxy power-spectrum predicted by the model with observations. Indeed, any sensible model should provide a reasonable agreement with the data since it should more or less follow the dark matter power-spectrum which is itself normalized to observations of large-scale clustering (through $\sigma_8$). Then, it seems difficult to obtain narrow constraints on the input parameters from only one curve, in view of the number of these free parameters and the approximations involved in such models (e.g., importance of the scatter of galaxy properties, shape of the distribution $P(N)$ within virialized halos, role of dark matter substructures,...).

Such approaches differ from ours in many respects. First, we use a mass function derived from a scaling ansatz for the dark matter density field (Balian \& Schaeffer 1989) which applies to the non-linear regime. It is fully specified by the sole function $H(x)$, obtained from counts-in-cells statistics, and by the behaviour of the non-linear power-spectrum. Most importantly, it allows us to get various multiplicity functions for different classes of objects in addition to ``just-virialized'' halos, defined by a relation $\Delta(M)$ or $R(M)$ which specifies their overall density contrast or radius as a function of their mass (Valageas \& Schaeffer 1997; Valageas et al. 2000). This enables us to bypass the link between galaxies and ``just-virialized'' halos which hinders the approaches described above. Indeed, we directly recognize the galaxies themselves, defined by a relation $R(M)$ obtained from cooling constraints (Valageas \& Schaeffer 1999, 2000). 

Then, as explained above the clustering of these objects is obtained (Bernardeau \& Schaeffer 1992,1999) from the behaviour of the $p-$point correlation functions $\xi_p$ which led to the mass function (\ref{etah}). Hence it does not require any additional information. Moreover, although we restrict ourselves to the two-point correlation functions, our formulation also predicts the $p-$point correlation functions of any order $p$ for all the objects we study here (galaxies, QSOs,...). This clearly illustrates the predictive power of our model. Besides, although we do not take into account the scatter of galaxy properties with respect to their mass, our procedure automatically includes the part of the scatter of the galaxy distribution with respect to the distribution of ``just-virialized'' halos which is due to the statistics of the dark matter itself. 

Finally, the spirit of our present study is rather different from the works which are mainly intended to describe galaxy clustering (Seljak 2000; Peacock \& Smith 2000) as we merely compare with observations the predictions of a general model of structure formation which was fully constrained in previous papers which dealt with its consequences for galaxies, QSOs, Lyman-$\alpha$ clouds and clusters. In particular, we have no free parameters left to improve the agreement with the observed galaxy clustering and we check our model against a vast array of observations, ranging from Lyman-$\alpha$ clouds up to clusters, through galaxies and QSOs. Thus, our goal is not to use observations of galaxy clustering to constrain some free parameters, but to check whether a sensible scenario of structure formation can match this data. Besides, since we use an analytic model where the various physical processes can be clearly identified, we can get some valuable insight into the clustering properties of the range of objects we study here. For instance, we shall see that a model like ours can explain in simple physical terms the qualitative and quantitative differences between the redshift evolution of the clustering of bright galaxies and faint galaxies, small Lyman$-\alpha$ forest clouds and damped systems, or QSOs and galaxies.

Another method to study the clustering of galaxies is to use N-body simulations, which may be supplemented by semi-analytic modelling of star-formation or galaxy formation (e.g., Benson et al. 2000a). In principle, this allows one to avoid the assumptions and approximations involved in an analytic model like ours to describe the dark matter density field. However, this technique may be limited by numerical resolution and computer time constraints. In any case, this is certainly a promising method but it does not provide yet the clustering properties of the full range of objects we study here in a unified fashion.

\section{Critical density universe}
\label{Critical universe}

We first consider the case of a critical density universe $\Om=1$ with a
CDM power-spectrum (Davis et al. 1985), normalized to
$\sigma_8=0.5$. We choose a baryonic density parameter
$\Omega_b=0.04$ and $H_0=60$ km/s/Mpc. We use the scaling function
obtained by Bouchet et al. (1991) for $H(x)$.

\subsection{Galaxies}
\label{Galaxies}

\subsubsection{Present universe}
\label{Present universe}

\begin{figure}[htb]

\centerline{\epsfxsize=8 cm \epsfysize=5.5 cm \epsfbox{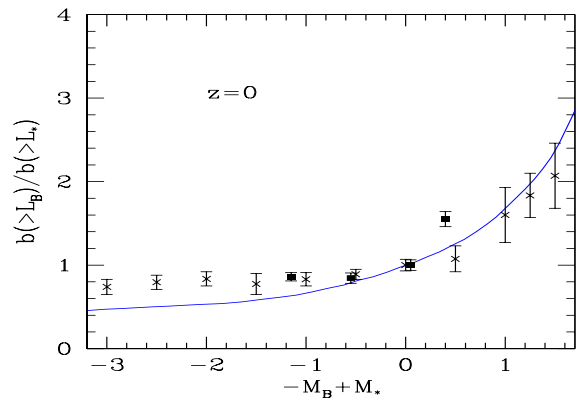}}

\caption{The luminosity dependence of the bias $b(>L_B)$ at $z=0$. The 
solid line is the mean bias $b(>L_B)/b(>L_*)$ of galaxies brighter
than $L_B$ normalized to the bias of galaxies brighter than $M_* =
-19.5$. The data points are from Benoist et al. (1996) (crosses) and
Willmer et al. (1998) (squares).}
\label{figbgalBenoistO1}

\end{figure}

We use the model of galaxy formation and evolution described in detail
in Valageas \& Schaeffer (1999), where it was checked against many
observations (luminosity function, Tully-Fisher relation...). We
define galaxies by two constraints: 1) {\it a virialization condition}
$\Delta>\Delta_c$ where $\Delta_c \sim 177$ is given by the usual
spherical model, and 2) {\it a cooling constraint} $t_{cool} < t_H$
which requires the gas to cool within a few Hubble times at
formation. At high virial temperature and low redshift, the condition
2) is the most stringent and can be approximated by a constant radius
constraint $R=R_{cool}$. At high redshift it becomes irrelevant since
any halo which satisfies 1) also satisfies 2). These two conditions
define for galaxies the constraint $\Delta(M,z)$ we introduced in
Sect.\ref{Multiplicity functions}. Once we have identified in this
manner the dark-matter halos which correspond to galaxies we still
need to set up a model for star formation to obtain the stellar
properties of these objects. As detailed in Valageas \& Schaeffer
(1999) our model involves 4 components: (1) short-lived stars which
are recycled, (2) long-lived stars which are not recycled, (3) a
central gaseous component which is deplenished by star formation and
ejection by supernovae winds, replenished by infall from (4), a diffuse
gaseous component. This allows us to obtain the luminosity of these
galaxies. Since observations usually refer to the mean bias (or
correlation function) of galaxies above a given luminosity, we define:
\beq 
b(>L) = \frac{ \int_L^{\infty} b(L) \; \eta_g(L) \;
\frac{dL}{L} } { \int_L^{\infty} \eta_g(L) \; \frac{dL}{L} } \; ,
\label{bg>L} 
\eeq
where we used the factorization of the bias, see (\ref{bx}). We
present in Fig.\ref{figbgalBenoistO1} the dependence of the bias on
the B-band luminosity.

We see that we recover the trend of larger bias for brighter galaxies,
although our dependence is somewhat steeper than the observed
relation. However, we note that in our analytical model each
luminosity is associated with a specific object of a given mass,
radius,... In a more realistic model one should include some scatter
in this relation which would lessen the final luminosity dependence of
the bias (for a large scatter such that the luminosity is uncorrelated
with the properties of the underlying dark matter halo we would have
no B-magnitude dependence left).

\begin{figure}[htb]

\centerline{\epsfxsize=8 cm \epsfysize=5.5 cm \epsfbox{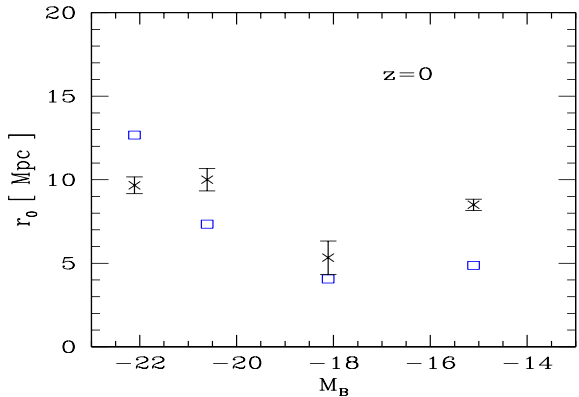}}

\caption{From left to right the correlation length $r_0$ at $z=0$ of the
correlation function of galaxies between the magnitude bins
$-22.8<M_B<-20.8$, $-20.8<M_B<-19.8$ and $-19.8<M_B<-15.8$ with all
galaxies (i.e. with $-22.8<M_B<-15.8$). The point on the far right is
the correlation length of all galaxies with $-22.8<M_B<-15.8$. The
data points (crosses) are from Loveday et al. (1995) and the squares
are our predictions.}
\label{figr0crossO1}

\end{figure}

From the relation (\ref{bx}) we can also obtain the cross-correlation
of galaxies of luminosity $L_B$ with all galaxies, in a fashion
similar to (\ref{bg>L}). We can convert the bias $b$ evaluated in this
manner into a correlation length $r_0(b)$ defined by:
\beq
r_0(b) = b^{2/\gam} \; r_0 \; ,
\eeq
where $\gam=1.8$ is the approximate slope of the correlation
function and $r_0$ is the correlation length of the dark matter density field. 
Indeed in our model the correlation functions of
various objects and of the matter distribution all have the same slope
in the non-linear regime (note that Dave et al. 1999 find from numerical 
simulations that the bias shows very little scale dependence, if any, in the 
non-linear regime). This prescription allows us to compare in
Fig.\ref{figr0crossO1} our results to observations performed by
Loveday et al. (1995).

We see that our results are compatible with the
observations. We find an  increase of the bias
(hence of $r_0$) for brighter galaxies in accordance with  the data, 
except for the last point at $-22.8<M_B<-20.8$ where 
 the bias is measured to be  slightly smaller
than for
$-20.8<M_B<-19.8$. Nevertheless, since this contradicts the trend
observed by Benoist et al. (1996) this disagreement is probably not
conclusive.

\begin{figure}[htb]

\centerline{\epsfxsize=8 cm \epsfysize=5.5 cm \epsfbox{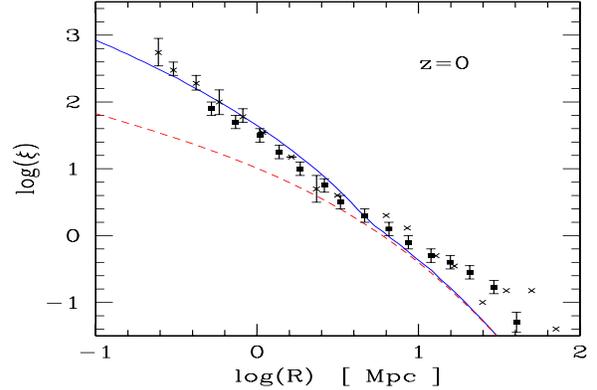}}

\caption{The correlation function of galaxies brighter than
$M_B=-18.6$ at $z=0$ (solid line). The dashed line is the linear
prediction. The data points are from Baugh (1996) (squares) for
$-21.6<M_B<-18.6$ and from Loveday et al. (1995) (crosses) for
$-22.8<M_B<-15.8$.}
\label{figXigalO1}

\end{figure}

We present in Fig.\ref{figXigalO1} the correlation function of
galaxies brighter than $M_B=-18.6$ at $z=0$. Our results agree
reasonably well with observations, although the slope we obtain may be
a bit too steep at large scales. This is a well-known discrepancy for CDM 
initial conditions with $\Om = 1$ and could be improved with
a slightly different initial power-spectrum. The dashed line is the
correlation function given by the linear theory multiplied by the
relevant bias factor. Hence it coincides with the actual correlation
function at large scales. At small scales the non-linear dynamics
increases the clustering as compared to the linear predictions, in
agreement with  observation. 
Note that in principle the non-linear predictions are valid for $\xia \gg 1$, 
but are in practice seen to hold as soon as $\xia \ge 1$.

\subsubsection{Redshift evolution}
\label{Redshift evolution}

\begin{figure}[htb]

\centerline{\epsfxsize=8 cm \epsfysize=5.5 cm \epsfbox{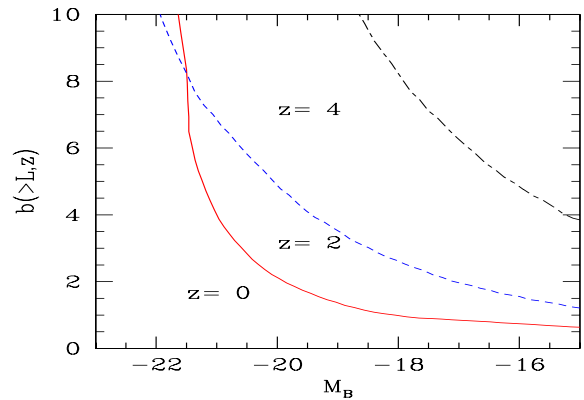}}

\caption{The redshift evolution of the relation luminosity-bias. The 
solid line corresponds to $z=0$, the dashed line to $z=2$ and the
dot-dashed line to $z=4$.}
\label{figbgal3zO1}

\end{figure}

We now turn to our predictions for the redshift evolution of the bias
associated with galaxies recognized by their B-band luminosity, shown
in Fig.\ref{figbgal3zO1}.

At ``low'' luminosities $M_B>-21$, the bias grows with redshift since
galactic halos correspond to increasingly rare overdense regions
(because structure formation was not as developed as it is
today). For very bright galaxies $M_B<-21$ we recover the same trend
except at low redshifts where the bias suddenly increases with
time. This evolution is due to a luminosity effect. Indeed, in our
model at low $z<1$ very massive galaxies see their luminosity decrease
with time as their initial gas content gets exhausted by
star formation (which is very efficient in this high-density
environment). As a consequence, when selecting galaxies by a fixed
B-band luminosity, one looks at more massive, higher virial temperature
halos at $z=0$ than at $z=1$. This shift to higher density
fluctuations acts opposite to the usual trend seen for lower
luminosity galaxies and leads to the increase with time of the bias at low $z$
for these very bright magnitudes.

\begin{figure}[htb]

\centerline{\epsfxsize=8 cm \epsfysize=5.5 cm \epsfbox{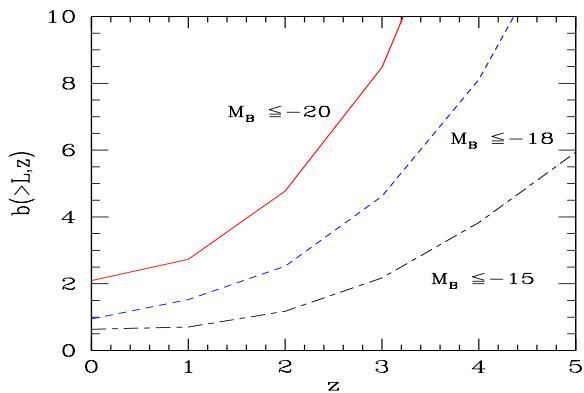}}

\caption{The redshift evolution of the bias associated with galaxies
below various B-band magnitude thresholds. The solid line corresponds
to $M_B \leq -20$, the dashed line to $M_B \leq -18$ and the
dot-dashed line to $M_B \leq -15$.}
\label{figb3MagO1}

\end{figure}

We present directly in Fig.\ref{figb3MagO1} the redshift evolution of
the bias associated with galaxies brighter than three B-band magnitude
thresholds. This clearly shows the dependence on luminosity of the
redshift evolution itself.

\begin{figure}[htb]

\centerline{\epsfxsize=8 cm \epsfysize=5.5 cm \epsfbox{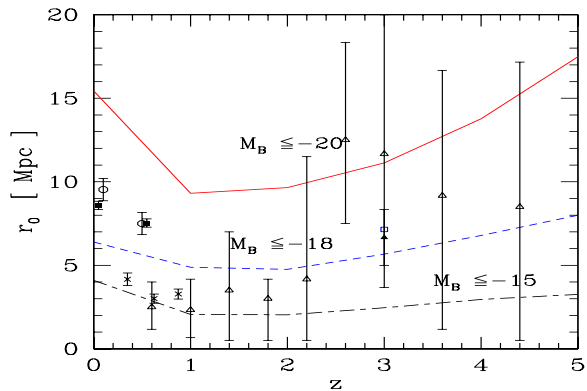}}

\caption{The redshift evolution of the correlation length $r_0$ (in
comoving Mpc) associated with galaxies below various B-band magnitude
thresholds. The solid line corresponds to $M_B \leq -20$, the dashed
line to $M_B \leq -18$ and the dot-dashed line to $M_B \leq -15$. The
data points are from Le Fevre et al. (1996) (crosses, $-21.3<M_B<-18.3$),
Carlberg et al. (1998) (filled squares, $M_B<-19.5$), Postman et al. (1998) 
(circles, $I<20$) and Magliocchetti \& Maddox (1999) (open triangles).
The filled triangle at $z=3$ corresponds to Lyman-break galaxies (observations 
from Adelberger et al. 1998) while the open square is our prediction for galaxies which have the same abundance as Lyman-break galaxies.}
\label{figr03MagO1}

\end{figure}

We show in Fig.\ref{figr03MagO1} the redshift evolution of the
comoving correlation length $r_0$ associated with galaxies brighter
than various magnitude thresholds.  The two data points at $z<0.1$
correspond to $z=0$ but they have been slightly shifted to be more
clearly seen. We can see that our results agree reasonably well with
observations, except for the data points from Le Fevre et al. (1996)
which however are inconsistent with data from Carlberg et al. (1998) and
Postman et al. (1998). The comoving number density of Lyman-break
galaxies at $z=3$ is $n \ga 2.7 \times 10^{-3}$ Mpc$^{-3}$ (see Adelberger
et al. 1998) which is also the abundance of galaxies brighter than
$M_B=-18.7$ (and $M \geq 5 \times 10^{11} \; M_{\odot}$) in our model. 
The square at $z=3$ shows our prediction for the clustering strength of galaxies
with this abundance. We see that this simple identification leads to good 
agreement with observations (filled triangle). The comoving correlation length 
of galaxies first decreases
at higher $z$, following the decline of the dark-matter two-point
correlation function. This trend is even larger for very bright
galaxies since it is amplified by the decline of their bias, as shown
in Fig.\ref{figb3MagO1}. However, at high redshifts $z>2$, $r_0$
increases with $z$ because of the strong growth of the bias. Of
course, this trend is stronger for brighter galaxies. We note that
contrary to Kauffmann et al. (1999) we obtain a ``dip'' in the
evolution of $r_0$, while these authors only found this feature for a
low-density flat cosmology. This is explained by the slower growth
with $z$ of the bias associated with galaxies in our model, as
compared to theirs, which is insufficient to override at low $z$ the
decrease of clustering measured by the decline of the dark-matter
correlation function.  Thus, such a feature depends on the
astrophysical model used to build galaxies.  However, we note that
observations show that $r_0$ decreases with larger $z$ in the range $0
\leq z \leq 0.5$, in agreement with our model. We recall here that our
results are direct consequences of a detailed model for structure
formation which has already been compared with many observations, so
that no attempt was made to ``improve'' the agreement of our
predictions with the data. In fact, as shown by the discrepancy
between various observations, the latter still have a significant
uncertainty. However, we recover the general trend of a decrease of $r_0$ up to 
$z \sim 2$ and a subsequent growth at higher redshifts. This behaviour also 
agrees with the results obtained by Baugh et al. (1999) from a semi-analytic 
model.

\subsection{Quasars}
\label{Quasars}

\begin{figure}[htb]

\centerline{\epsfxsize=8 cm \epsfysize=5.5 cm \epsfbox{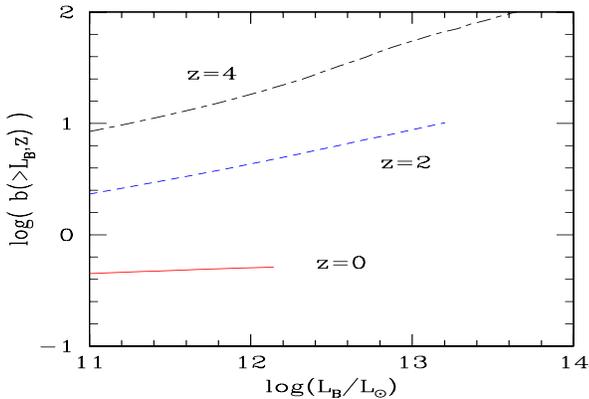}}

\caption{The redshift evolution of the bias associated with quasars above 
various luminosity thresholds. The solid line corresponds to
$z=0$, the dashed line to $z=2$ and the dot-dashed line to $z=4$.}
\label{figbQuas4zO1}

\end{figure}

From the description of galaxies presented above we have also
developed a model for the quasar luminosity function, as detailed in
Valageas \& Silk (1999a). We assume that a fraction $\lambda_Q \leq 1$
of galactic halos host a quasar with a mass $M_Q$ proportional to the
mass of gas $M_{gc}$ available in the inner parts of the galaxy: $M_Q
= F \; M_{gc}$. For most galaxies this also implies $M_Q \simeq F \;
M_s$ where $M_s$ is the stellar mass and we use $F=0.01$. We also assume
that quasars shine at the Eddington limit with a radiative efficiency
$\epsilon = 0.1$, which means that the quasar life-time is $t_Q = 4.4
\; \epsilon \; 10^8$ yr. This determines the quasar bolometric
luminosity:
\beq
L_Q = \frac{ \epsilon M_Q c^2}{t_Q} \; .
\label{LQ}
\eeq
We present in Fig.\ref{figbQuas4zO1} the redshift evolution of the
quasar luminosity - bias relation. Here $b(>L_B,z)$ is the bias of
quasars brighter than $L_B$ at redshift $z$.

We see that the bias increases with redshift since at higher $z$
quasars of a given luminosity correspond to rarer density
fluctuations. The high luminosity cutoff seen in the figure comes from
the fact that in our model very massive galaxies have already consumed
most of their gas content at low $z$, as explained in the previous
section, so that the maximum quasar luminosity declines with time
because of fuel exhaustion. Of course, at higher $z$ the quasar
luminosity function cutoff also declines because of the knee of the
mass function of collapsed objects (Valageas \& Silk 1999a).

\begin{figure}[htb]

\centerline{\epsfxsize=8 cm \epsfysize=5.5 cm \epsfbox{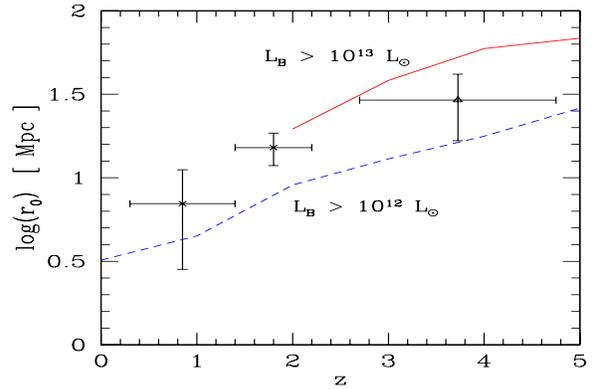}}

\caption{The redshift evolution of the comoving correlation length associated 
with quasars above 
various luminosity thresholds. The dashed line corresponds to $L_B>10^{12} \; 
L_{\odot}$ and the solid line to $L_B>10^{13} \; L_{\odot}$. The two crosses are 
observations from La Franca et al. (1998) for $0.3<z<1.4$ (lower point) and 
$1.4<z<2.2$ (upper point). The triangle shows the result at $2.7<z<4.75$ of 
observations by Stephens et al. (1997).}
\label{figr0Quas4zO1}

\end{figure}

From the bias $b(>L_B,z)$ we can derive the comoving correlation length of 
quasars. This is shown in Fig.\ref{figr0Quas4zO1}. We can check that our results agree reasonably well with observations. In particular, we recover the increase 
with redshift of the correlation length. La Franca et al. (1998) noticed that 
this behaviour differs from the decrease of $r_0$ observed for galaxies from 
$z=0$ to $z=1$ which we also obtain, as shown in Fig.\ref{figr03MagO1}. Thus a 
simple hierarchical model like ours is able to simultaneously explain both of 
these redshift dependences. The stronger increase with redshift of the 
correlation length associated with quasars, as compared with galaxies, also 
translates into the faster growth of their bias factor, see 
Fig.\ref{figbgal3zO1} and Fig.\ref{figbQuas4zO1}. This difference between 
galaxies and quasars is due to the redshift evolution of their respective ratios of (mass)/(luminosity). As described above, in our model the luminosity of a quasar  is proportional to its mass of gas, see (\ref{LQ}), and to the mass of stars in 
the host galaxy. Hence the bias of quasars of a given luminosity grows strongly 
with redshift since objects with a fixed mass of central gas are increasingly 
rare at earlier times when structure formation is less advanced. This overrides 
the decrease of the dark matter correlation length with $z$ and it leads to a 
rise of $r_0(>L_B)$, as seen in Fig.\ref{figr0Quas4zO1}. On the other hand, 
galaxies with a constant mass of star-forming gas had a higher luminosity in the 
past because the ratio $M_s/L$ decreases with redshift (see for instance Fig.17 
in Valageas \& Schaeffer 1999). This is due to the fact that the smaller age of 
the universe $t_H$ selects more massive and brighter stars with a life-time $t_* 
\propto t_H$. In other words, at later times the contribution from faint 
long-lived stars which progressively accumulated in the galaxy ever since it was 
born becomes increasingly important as compared with the contribution of bright 
newly-born short-lived stars which quickly disappear as supernovae and thus 
cannot accumulate. This implies that by looking at a fixed luminosity one 
selects a smaller mass of gas and stars at higher redshift. 

Thus, if we start at $z=0$ with galaxy and quasar luminosities which correspond 
to the same host halos, at large redshift $z>0$ the galaxies with the same 
luminosity correspond to smaller halos than the quasar hosts. Hence galaxies 
correspond to less extreme objects and have a smaller bias. Thus at low $z$ 
their correlation length follows the decreases of the dark matter $r_0$, until 
$z \sim 1$, beyond which the rise of their bias becomes sufficiently strong to 
lead to a growth of $r_0(>L_B)$, see Fig.\ref{figr03MagO1}. The precise 
amplitude of these redshift evolutions depends somewhat on the astrophysical 
models used for quasars and stars (IMF,...). However, the agreement we obtain 
with these observations is remarkable since our model was built independently of 
these considerations, with no free parameter left for this present discussion. Thus this 
behaviour appears as a natural outcome of such models which suggests they 
provide a realistic description of the global properties of star and quasar 
formation, even though the details of these processes are still poorly known.

\subsection{Lyman$-\alpha$ absorbers}
\label{Lyman-alpha absorbers}

As shown in Valageas et al. (1999), the formalism introduced in
Sect.\ref{Multiplicity functions} also allows us to build a model
for Lyman-$\alpha$ clouds, some of which are objects with a lower
density than the mean density of the universe. We describe these
absorbers as three different populations of objects. Low density halos
with a small virial temperature see their baryonic density
fluctuations erased over scales of the order of the Jeans length
$R_J$. These mass condensations form our first class of objects,
defined by the scale $R_J$, which can be identified with the
Lyman-$\alpha$ forest. They correspond to low density clouds which can
even be underdense (but stick out within very low-density regions or
voids). On the other hand, potential wells with a large virial
temperature maintain their density profile. Thus, they constitute a
second population of absorbers where the column density observed along
a line of sight depends strongly on the impact parameter. They can be
identified with the Lyman-limit systems. Finally, the deep neutral
cores of these halos (because of self-shielding) form our third class
of objects which correspond to the damped systems.

\begin{figure}[htb]

\centerline{\epsfxsize=8 cm \epsfysize=5.5 cm \epsfbox{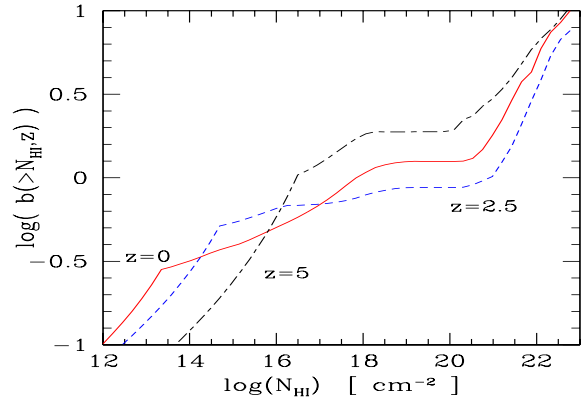}}

\caption{The redshift evolution of the bias associated with Lyman-$\alpha$ 
absorbers above various column density thresholds. The solid line
corresponds to $z=0$, the dashed line to $z=2.5$ and the dot-dashed
line to $z=5$.}
\label{figblym3z}

\end{figure}

We present in Fig.\ref{figblym3z} the redshift evolution of the bias
associated with Lyman-$\alpha$ absorbers above various column density
thresholds.

\begin{figure}[htb]

\centerline{\epsfxsize=8 cm \epsfysize=5.5 cm \epsfbox{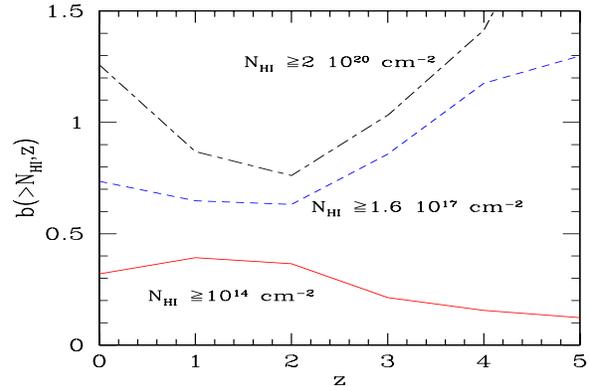}}

\caption{The redshift evolution of the bias associated with Lyman-$\alpha$ 
absorbers above various column density thresholds. The solid line
corresponds to $N_{HI} \geq 10^{14}$ cm$^{-2}$, the dashed line to
$N_{HI} \geq 1.6 \; 10^{17}$ cm$^{-2}$ and the dot-dashed line to
$N_{HI} \geq 2 \; 10^{20}$ cm$^{-2}$.}
\label{figb3NHIO1}

\end{figure}

We note that large column density absorbers have a bias greater than
unity since they correspond to high density fluctuations (massive
halos) while forest clouds which are low density fluctuations located
in underdense areas have a small bias $b<1$. For $N_{HI}<10^{14}$
cm$^{-2}$ the bias decreases at larger redshift because the higher
mean density of the universe implies that to observe a constant small
column density one must look at increasingly underdense objects. In a
similar fashion, for Lyman-limit or damped systems the bias first
declines from $z=0$ to $z=2.5$. However, since these objects
correspond to virialized halos the bias eventually grows with $z$ as
the structure formation process is less advanced. Of course there is
also an intermediate range of $N_{HI}$ where the behaviour is more
intricate. We display the redshift evolution of the bias associated
with clouds above three different column densities thresholds in
Fig.\ref{figb3NHIO1}.

We see that the redshift evolution of the bias of Lyman-$\alpha$
absorbers can display a non-trivial behaviour, since many processes
play a role in the properties of these clouds, including the amplitude
of the UV background radiation field. Thus, a phenomenological
parameterisation of the form $b \propto N_{HI}^{\alpha} \;
(1+z)^{\beta}$ is unlikely to provide an accurate description. Note
however that intermediate clouds in the range $10^{16} < N_{HI} <
10^{21}$ cm$^{-2}$ have a bias which remains reasonably close to unity
at all redshifts $z<5$ and thus can be used as tracers of the matter
density field. A comparison of our results with observations for the two-point 
correlation function in velocity space is described in Valageas et al. (1999).

\subsection{Clusters}
\label{Clusters}

\begin{figure}[htb]

\centerline{\epsfxsize=8 cm \epsfysize=5.5 cm \epsfbox{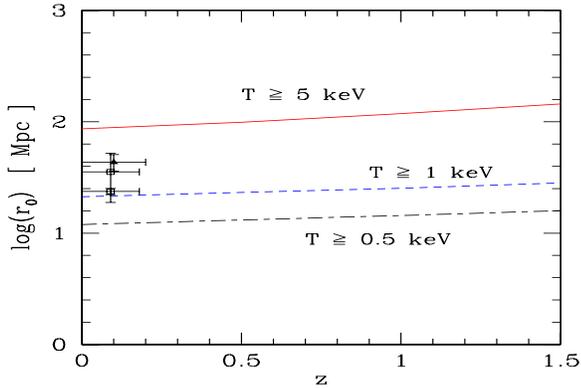}}

\caption{The redshift evolution of the comoving correlation length associated 
with clusters above 
various virial temperature thresholds. The dot-dash line corresponds to $T \geq 
0.5$ keV, the dashed line to $T \geq 1$ keV and the solid line to $T \geq 5$ 
keV. The two squares are observations from Croft et al. (1997) at $z<0.18$ for 
clusters with a richness ${\cal R} \geq 50$ (lower point) and ${\cal R} \geq 
110$ (upper point). The triangle shows the result at $z<0.2$ of observations by 
Borgani et al. (1999) for the X-ray brightest Abell cluster sample.}
\label{figr0clusO1}

\end{figure}

Finally, we show in Fig.\ref{figr0clusO1} the redshift evolution of the 
correlation length associated with clusters above various virial temperature 
thresholds. Here we simply defined clusters as ``just-virialized'' halos 
characterized by the usual density contrast $\Delta_c \sim 177$. A detailed 
description of our model for clusters and the evolution of their X-ray 
luminosity function is presented in Valageas \& Schaeffer (2000). We see that 
our results agree reasonably well with observations. As was the case for 
quasars, the rise with redshift of the bias associated to these halos leads to 
an increase of their correlation length. However, this increase is slower than 
for quasars, see Fig.\ref{figr0Quas4zO1}, because here we select these halos by 
a constant temperature threshold. This implies that their mass evolves as $M 
\propto (1+z)^{-3/2}$ hence we consider smaller halos at high $z$ which are more 
common (and have a smaller bias) than objects of constant mass. Thus, we have 
obtained a simple model of structure formation which is consistent with 
observations from Lyman-$\alpha$ clouds ($M \sim 10^8 \; M_{\odot}$) up to 
clusters ($M \sim 10^{15} \; M_{\odot}$), i.e. over 7 orders of magnitude.

\subsection{Distribution of matter}
\label{Distribution of matter}

The models we have built in previous studies for galaxies (Valageas \&
Schaeffer 1999) and Lyman-$\alpha$ clouds (Valageas et al. 1999) allow
us to obtain the redshift evolution of the fraction of baryonic matter
associated with various components.

\begin{figure}[htb]

\centerline{\epsfxsize=8 cm \epsfysize=5.5 cm \epsfbox{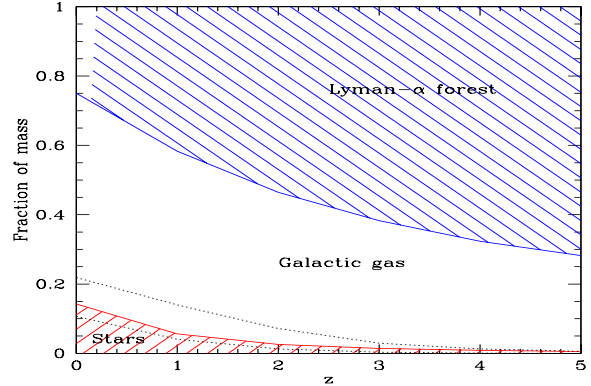}}

\caption{The redshift evolution of the fraction of baryonic mass embedded 
within Lyman-$\alpha$ forest clouds (upper shaded area), galactic
halos (white area) and stars (lower shaded area). The dotted lines
separate the contributions from bright ($M_B<-19.5$, lower part) and
faint ($M_B>-19.5$, upper part) galaxies for stars and gas, respectively.}
\label{figfracO1}

\end{figure}

We present in Fig.\ref{figfracO1} the contributions of stars, galactic
gas and Lyman-$\alpha$ forest clouds. In our model the sum of these
three constituents is unity. Indeed, all the matter is embedded within
mass condensations which we identify with galaxies (deep potential
wells where the gas can cool) or with Lyman-$\alpha$ forest clouds
(small density fluctuations heated by the UV background radiation). We
see that the mass fraction within stars is always small ($\sim 10 \%$
at $z=0$) while the contribution of Lyman-$\alpha$ forest clouds shows
a fast increase with redshift to constitute $\sim 70 \%$ of the matter
at $z=5$ when there are few deep potential wells. Bright galaxies
($M_B<-19.5$) contain most of the mass embedded within stars but they
only constitute a small part of the gas which was able to cool. Moreover, their 
total mass of gas starts declining at low
redshifts $z<1$ as it is converted into stars. This trend does not
appear for faint galaxies because their star formation process is less
efficient while they accrete gas which was previously part of the
Lyman-$\alpha$ forest clouds. Moreover, some of these small galaxies have just formed as new potential wells have built up. Note that if the IGM is 
reheated by supernovae or quasars up to $T \sim 5 \; 10^5$ K, which is suggested by the observed relation temperature - X-ray luminosity of clusters (e.g., Ponman et al. 1999), the fraction of matter which has been able to cool and appears as galactic gas in Fig.\ref{figfracO1} would be smaller. This was studied in details in Valageas \& Silk (1999b).

\begin{figure}[htb]

\centerline{\epsfxsize=8 cm \epsfysize=5.5 cm \epsfbox{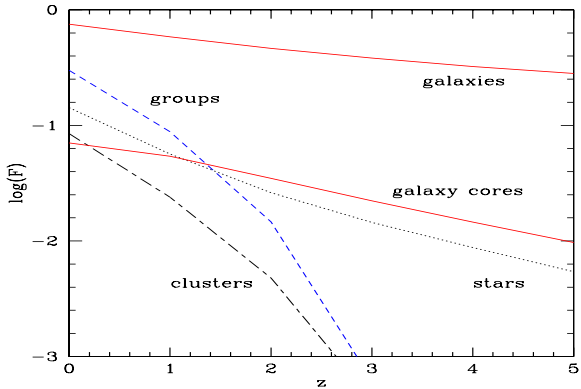}}

\caption{The redshift evolution of the fraction $F$ of baryonic mass
embedded within galaxies (upper solid line: total mass, lower solid
line: gas in the core), groups (dashed line), clusters (dot-dashed
line) and stars (dotted line).}
\label{figclusO1}

\end{figure}

\begin{table}
\begin{center}
\caption{Baryonic parameter $\Omega_i$ associated with various objects at 
$z=0$ and $z=3$, for a critical universe with $\Omega_b=0.04$. The
second line which appears for some contributions is the observational
estimate from Fukugita et al. (1998).}
\label{table1}
\begin{tabular}{ccc}\hline

  & $z=0$ & $z=3$ \\
\hline\hline

stars & 0.0057 & 0.0006 \\
 & 0.002 - 0.0057 & \\
\hline

galactic gas & 0.024 & 0.015 \\
\hline

groups & 0.012 & 0.00002 \\
 & 0.011 - 0.045 & \\
\hline

clusters & 0.0034 & 0.00002 \\
 & 0.0014 - 0.0044 & \\
\hline

Lyman-$\alpha$ forest & 0.01 & 0.025 \\
  & & 0.01 - 0.05 \\
\hline

\end{tabular}
\end{center}
\end{table}

We show in Fig.\ref{figclusO1} the mass fraction we associate with
groups and clusters of galaxies, as well as with stars and
galaxies. We define groups as virialized halos (i.e. with a density
contrast $\Delta_c$) which do not satisfy the constraint $t_{cool} <
t_H$. They correspond to massive potential wells with a large virial
temperature (at high $T$ we have $t_{cool} \sim \sqrt{T}$). At high
redshifts this contribution declines since an increasing proportion of
virialized halos can cool (indeed $t_{cool} \propto 1/\rho$ while $t_H
\propto 1/\sqrt{\rho}$), see also Valageas \& Schaeffer (1999). We
identify clusters as groups with a virial temperature $T \geq 1$
keV. At low $z$ they only form a small sub-class of galaxy groups
while at high $z$ all groups have large virial temperatures $T \geq 1$
keV as explained above. Note that in our model the mass associated
with groups is also associated to the member galaxies, so that it is a
part of the fraction of matter recognized as embedded within galaxies,
shown in Fig.\ref{figfracO1} and in Fig.\ref{figclusO1} (upper solid
line). A detailed study of groups and clusters in presented in Valageas \& 
Schaeffer (2000).

From the fraction of matter $F_i$ associated with various objects and
the baryonic density parameter $\Omega_b=0.04$ we used in our model, we
also obtain the baryonic density parameters $\Omega_i= F_i \Omega_b$ for various classes of objects. We display in Tab.\ref{table1}
a comparison of our results with observational estimates by Fukugita
et al. (1998). We note that our predictions are consistent with the
data. In Tab.\ref{table1} we removed from the mass embedded within
clusters and groups shown in Fig.\ref{figclusO1} the fraction of
baryons in stars or galactic cores which they contain.

\section{Open universe}
\label{Open universe}

Using the same formalism we can also study the case of an open universe 
$\Om=0.3$, $\Ol=0$, with a CDM power spectrum normalized to
$\sigma_8=0.77$. We use $\Omega_b=0.03$ and $H_0=60$ km/s/Mpc.

\subsection{Galaxies}
\label{Galaxiesopen}

\begin{figure}[htb]

\centerline{\epsfxsize=8 cm \epsfysize=5.5 cm \epsfbox{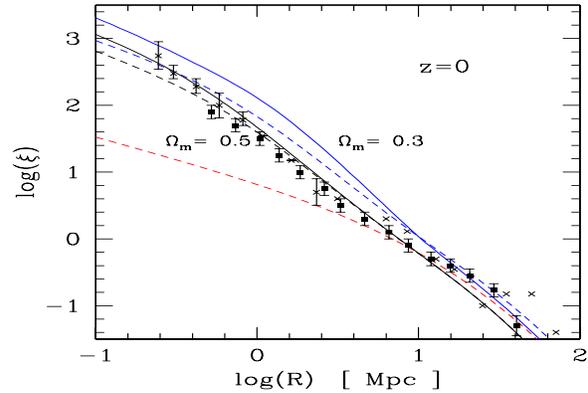}}

\caption{The correlation function of galaxies brighter than
$M_B=-18.6$ at $z=0$ as in Fig.\ref{figXigalO1}. The upper solid line corresponds to the open universe with $\Om=0.3$ and $\Ol=0$, while the lower dashed-line is the linear prediction. The upper dashed line shows the low-density flat universe with $\Om=0.3$ and $\Ol=0.7$. The middle solid line (resp. dashed line) displays the case $\Om=0.5$ and $\Ol=0$ (resp. $\Om=0.5$ and $\Ol=0.5$).}
\label{figXigalO03}

\end{figure}

We show in Fig.\ref{figXigalO03} the correlation function of galaxies
brighter than $M_B=-18.6$ at $z=0$ (upper solid line). Our predictions are similar to the
results obtained in Fig.\ref{figXigalO1}. However, a bump appears at 
intermediate scales ($R \sim 2$ Mpc) where $\xia \sim 10$. This corresponds to 
the sharp increase of $\xia(R)$ from the linear regime ($\xia \la 1$) to the 
non-linear regime ($\xia \ga 100$). This leads to a ``step-like'' profile which 
is also seen in numerical simulations (e.g., Valageas et al. 2000) and which gets 
larger for low density universes (Peacock \& Dodds 1996). This is due to the 
slow-down of the linear growth factor while the two-point correlation function 
keeps increasing as $\xia(R,a) \propto a^3$ in the highly non-linear regime 
(Valageas \& Schaeffer 1997). 
The observed two-point correlation function on the other hand is surprisingly 
close to a power-law as soon as $\xia \ge 1$.
Such a behaviour requires a precise balance between the curvature of the initial 
power-spectrum, the distortions induced by the non-linear dynamics and the bias 
of galaxies (which also depends on non-gravitational processes which affect 
galaxy formation). 
Note however that using numerical simulations Pearce et al. (1999) and Benson et al. (2000a) obtain a good match to observations as the galaxies they get are 
antibiased relative to the dark matter on small scales which ``conspires'' to 
provide a power-law galaxy correlation function.
 This behaviour is partly due to the spatial exclusion of halos. 
Moreover, the expression (\ref{bx1}) we use for the bias is only valid at large 
separations $r \gg R$ between objects of size $R$. Thus, the discrepancy at 
small scales might be cured by a more rigorous calculation which would not use 
the approximation $r \gg R$.

\begin{figure}[htb]

\centerline{\epsfxsize=8 cm \epsfysize=5.5 cm \epsfbox{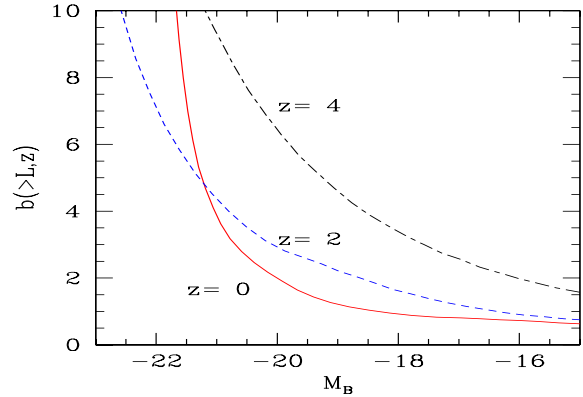}}

\caption{The redshift evolution of the relation luminosity-bias. The 
solid line corresponds to $z=0$, the dashed line to $z=2$ and the
dot-dashed line to $z=4$.}
\label{figbgal3zO03}

\end{figure}

On the other hand, note that in the case of a critical density universe this 
``bump'' in the mildly non-linear regime is much smaller, which leads to good 
agreement with observations (see Fig.\ref{figXigalO1}). In particular, we also display in Fig.\ref{figXigalO03} the dark matter correlation function we would obtain for a low-density flat universe (upper dashed line) with $\Om=0.3$ and $\Ol=0.7$, with a slightly different power-spectrum from Bardeen et al. (1989) (and $\Gamma=0.18$) using the non-linear fit from Peacock \& Dodds (1996). Thus this corresponds to a bias of unity, which is a good approximation for galaxies with $M_B \leq -18.6$ at $z=0$, see Fig.\ref{figbgalBenoistO1} and Fig.\ref{figbgal3zO03}. The main point is that the ``bump'' is much weaker for a flat cosmology, as can also be seen in Peacock \& Dodds (1996). The intermediate lines show the dark matter correlation functions obtained for $\Om=0.5$ (with $\Ol=0$ and $\Ol=0.5$) and normalized to $\sigma_8 = 0.6$ (and $\Gamma=0.3$). Thus, the shape itself of the galaxy correlation function does not allow one to strongly discriminate between these various cosmological scenarios. Although the ``bump'' which appears for the open universe with $\Om=0.3$ seems to disfavour this model, the inaccuracy involved in the modelling of the galaxy clustering properties is certainly too large to draw meaningful conclusions, especially since numerical simulations suggest that simple physical processes could remove this feature (Benson et al. 2000a).

We present in Fig.\ref{figbgal3zO03} the redshift evolution of the luminosity-bias relation. We can see that our results are again similar
to those obtained in Fig.\ref{figbgal3zO1}, since the astrophysical
model is the same and it must satisfy the same observational
constraints at low $z$. However, the bias of bright galaxies increases more 
slowly with redshift than for a critical density universe (compare the curve at 
$z=4$ with Fig.\ref{figbgal3zO1}). This is due to the higher normalization 
$\sigma_8$ of the power-spectrum and to the slower redshift evolution of the 
linear growth factor which means that massive galactic halos correspond to less 
extreme density fluctuations at $z=4$ than for the case $\Om=1$ when they 
have a similar clustering pattern at $z=0$.

\begin{figure}[htb]

\centerline{\epsfxsize=8 cm \epsfysize=5.5 cm \epsfbox{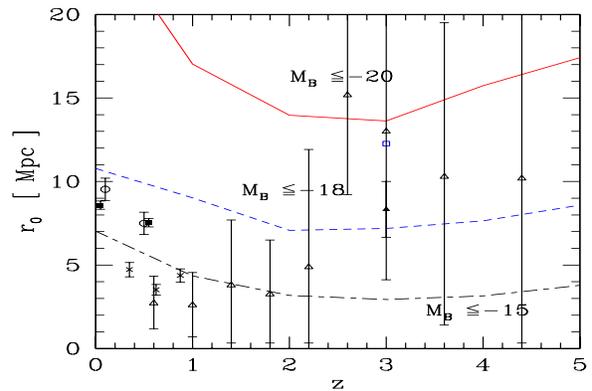}}

\caption{The redshift evolution of the correlation length $r_0$ (in
comoving Mpc) associated with galaxies below various B-band magnitude
thresholds, as in Fig.\ref{figr03MagO1}, for an open universe.}
\label{figr03MagO03}

\end{figure}

We show in Fig.\ref{figr03MagO03} the redshift evolution of the
comoving correlation length $r_0$ associated with galaxies below
various B-band magnitude thresholds. The galaxy correlation length is
larger than for the previous case of a critical density universe because the 
normalization of the power-spectrum
($\sigma_8$) is slightly higher. This also implies a slower redshift evolution. 
We note that the clustering pattern we predict for Lyman break galaxies is 
somewhat too strong as compared to observations. However, this could be due to 
some extinction by dust of the luminosity of these LBG. Indeed, in our model the 
number of galaxies brighter than $M_B=-18$ is $n=3 \times 10^{-3}$ Mpc$^{-3}$ while 
the abundance of LBG is $n \sim 8.5 \times 10^{-4}$ Mpc$^{-3}$ (Adelberger at 
al. 1998). This means that we would recover the right correlation length if we 
assumed that LBG correspond to $30 \%$ of these galaxies with $M_B \leq -18$. An 
alternative would be that LBG correspond to galaxies which are in the midst of a 
starburst phase (e.g., Somerville et al. 2000) which increases their luminosity 
(note that the absolute LBG magnitude is rather large: $M_R \leq -21.8$ for an 
apparent magnitude ${\cal R} \leq 25.5$ without K-correction). In this case, 
Lyman-break
galaxies would also be drawn from a larger parent population which could again 
lead to a smaller correlation length. Note on the other hand that it would be 
difficult to reconcile the model with observations if we had predicted a smaller 
value of $r_0$ than the one given by the data.

\begin{figure}[htb]

\centerline{\epsfxsize=8 cm \epsfysize=5.5 cm \epsfbox{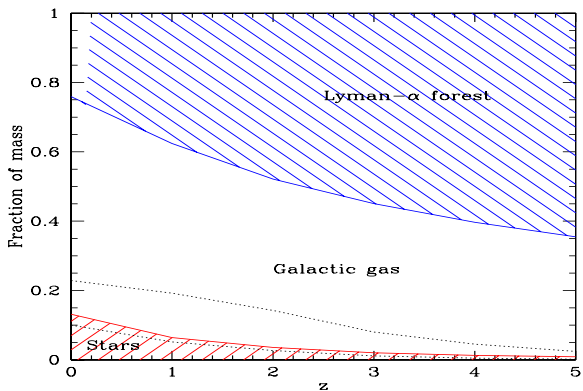}}

\caption{The redshift evolution of the fraction of baryonic mass embedded 
within Lyman-$\alpha$ forest clouds (upper shaded area), galactic
halos (white area) and stars (lower shaded area). The dotted lines
separate the contributions from bright ($M_B<-19.5$, lower part) and
faint ($M_B>-19.5$, upper part) galaxies for stars and gas.}
\label{figfracO03}

\end{figure}

\subsection{Distribution of matter}

We show in Fig.\ref{figfracO03} the contributions of stars, galactic
gas and Lyman-$\alpha$ clouds to the baryonic density in the universe
as a function of redshift. We obtain a behaviour similar to
Fig.\ref{figfracO1}.

\begin{figure}[htb]

\centerline{\epsfxsize=8 cm \epsfysize=5.5 cm \epsfbox{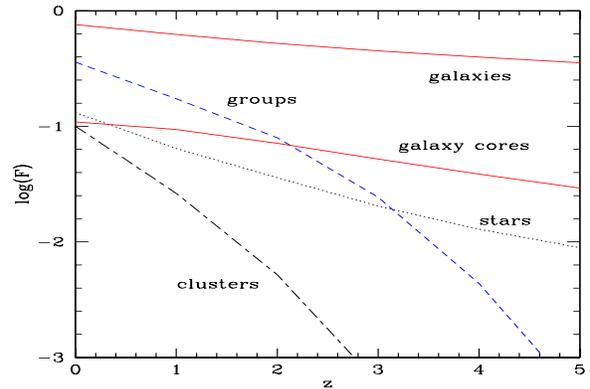}}

\caption{The redshift evolution of the fraction $F$ of baryonic mass
embedded within galaxies (upper solid line: total mass, lower solid
line: cores), groups (dashed line), clusters (dot-dashed line) and
stars (dotted line).}
\label{figclusO03}

\end{figure}

We display in Fig.\ref{figclusO03} the redshift evolution of the
fraction of matter embedded within galaxies, groups, clusters and
stars. We see that the mass fraction formed by groups and clusters declines
more slowly than in Fig.\ref{figclusO1}. Again this is due to the higher 
normalization $\sigma_8$ and to the slower redshift evolution of the linear 
growth factor.

\begin{table}
\begin{center}
\caption{Baryonic parameter $\Omega_i$ associated with various objects
at $z=0$ and $z=3$, for an open universe with $\Om=0.3$,
$\Omega_b=0.03$. The second line which appears for some contributions
is the observational estimate from Fukugita et al. (1998).}
\label{table2}
\begin{tabular}{ccc}\hline

  & $z=0$ & $z=3$ \\
\hline\hline

stars & 0.0039 & 0.0006 \\
 & 0.002 - 0.0057 & \\
\hline

galactic gas & 0.019 & 0.013 \\
\hline

groups & 0.011 & 0.0007 \\
 & 0.011 - 0.045 & \\
\hline

clusters & 0.003 & 0.00002 \\
 & 0.0012 - 0.0044 & \\
\hline

Lyman-$\alpha$ forest & 0.007 & 0.016 \\
  & & 0.01 - 0.05 \\
\hline

\end{tabular}
\end{center}
\end{table}

We show in Tab.\ref{table2} the baryonic density parameters $\Omega_i$
associated with various objects. We note that our results still agree
reasonably well with observations.

\section{Conclusion}

In this article we have described the redshift evolution of the
clustering of various objects (galaxies, quasars, Lyman-$\alpha$
clouds, clusters) within the framework of a unified model of structure formation
which has already been checked against many observations. Hence there
are no free parameters chosen to fit the observations considered in
this article, since the model was already overconstrained by
previous studies. We have shown that {\it the bias of these astrophysical
objects displays an intricate behaviour} which cannot be described by
simple power-law parameterizations. Indeed, the link between these
objects and the underlying dark matter component reflects the
influence of many factors (cooling processes, star formation
efficiency, evolution of the UV background radiation field...) which
can have competing effects. We have shown that in our model the bias
associated with galaxies strongly increases with their
luminosity. 
This was already emphasized
 in the earlier calculations
 (Schaeffer 1987; Bernardeau \& Schaeffer 1992) as
 being a genuine outcome of the hierarchical clustering hypothesis.
We recover the trend seen in observations, although
our dependence may be slightly too strong. However, this could be due to
the simplicity of our model which does not include any scatter in the
properties of the objects we model. The redshift evolution we obtain
also agrees with data, in particular the correlation length of
Lyman-break galaxies observed at $z = 3$ is consistent with our
results. 

We have also described the bias associated with Lyman-$\alpha$ clouds. 
Thus, it appears that absorbers in the range
$10^{16} < N_{HI} < 10^{21}$ cm$^{-2}$ have a bias close to unity
(hence they trace the dark matter density field) in contrast to other
clouds and galaxies which show a very different behaviour of their
clustering properties. Our results also agree with observations for quasars and 
clusters. In particular, they explain in a natural fashion the qualitative and quantitative differences between the redshift evolution of the clustering of QSOs and of galaxies. Thus, {\it our model provides a good description of the intrinsic properties and the spatial clustering of a wide variety of objects}, from small underdense Lyman-$\alpha$ absorbers up to massive clusters, {\it which covers six orders of magnitude in mass}.

Finally, we have described the redshift evolution of the distribution
of matter over the various objects observed in the universe. Our
results show a reasonable agreement with observations. This is not
surprising since our model has already been checked against the galaxy
luminosity function, the number density of Lyman-$\alpha$ clouds,
which constrain the amount of matter embedded within these objects. In
particular, it appears that {\it light does not trace baryonic mass} as
bright galaxies ($M_B<-19.5$) which contain most of the stars only
provide a small fraction of the mass associated with virialized and
cooled halos.

We have also performed the same analysis for the case of an
open universe. In both cosmologies our results agree reasonably well with 
observations but the details of astrophysical models are still too uncertain to
discriminate between these two cases. However, for the open universe the galaxy 
correlation function we obtain shows a shoulder which does not appear in 
observations. This could be due to the approximation involved in our calculation which is only valid on scales much larger than the size of the objects. In particular, some numerical simulations do not get this ``bump'' (Benson et al. 2000a). On the other hand, they do not recover the luminosity dependence of the galaxy bias (Benson et al. 2000b) and simulations cannot simultaneously describe the clustering of the whole range of objects we studied in this paper. Nevertheless, the fact that our
predictions are nearly  consistent with the data over such a wide range of
objects shows that hierarchical scenarios provide a very good
description of structure formation. On the other hand, more detailed
observations of the amplitude of the clustering of galaxies selected
by their luminosity will constrain the astrophysical models.

\begin{acknowledgements}

Much of this work was performed at the Center for Particle Astrophysics, University of California, Berkeley, whose hospitality we gratefully acknowledge.

\end{acknowledgements}

\end{document}